\documentclass[a4paper,11pt]{article}
\pdfoutput=1 

\usepackage{jheppub} 
\usepackage{dsdshorthand}
\usepackage{amsbsy}
\usepackage{caption}
\usepackage{subcaption}
\usepackage{fancyvrb}
\usepackage{graphicx}
\usepackage{xcolor}
\graphicspath{{.}}

\usepackage[T1]{fontenc} 

\title{\boldmath Non-saturation of Bootstrap Bounds by Hyperbolic Orbifolds}

\author{Alex Radcliffe}

\affiliation{King's College London,\\Strand, London, WC2R 2LS, UK}

\emailAdd{alexander.radcliffe@kcl.ac.uk}

\abstract{In recent years the conformal bootstrap has produced surprisingly tight bounds on many non-perturbative CFTs. It is an open question whether such bounds are indeed saturated by these CFTs. A toy version of this question appears in a recent application of the conformal bootstrap to hyperbolic orbifolds, where one finds bounds on Laplace eigenvalues that are exceptionally close to saturation by explicit orbifolds. In some instances, the bounds agree with the actual values to 11 significant digits. In this work we show, under reasonable assumptions about the convergence of numerics, that these bounds are not in fact saturated. In doing so, we find formulas for the OPE coefficients of hyperbolic orbifolds, using links between them and the Rankin-Cohen brackets of modular forms.}

\begin{document} 
	\maketitle
	\flushbottom
	\section{Introduction}
	The conformal bootstrap has been used to great effect in recent years to explore the space of conformal field theories (CFTs)~\cite{Simmons-Duffin:2016gjk, Poland:2018epd}, and to constrain scaling dimensions of operators in these theories. These techniques have also been adapted to constrain the spectra of the Laplace-Beltrami and Dirac operators on hyperbolic orbifolds~\cite{Bonifacio:2021msa, Kravchuk:2021, Bonifacio:2021, Bonifacio:2023ban, Gesteau:2023brw}. As well as being of great interest in pure mathematics, this provides a useful toy model through which to test bootstrap techniques.
	
	In particular, it was found in~\cite{Kravchuk:2021} and~\cite{Bonifacio:2021} that the spectral gap,\footnote{The smallest non-zero eigenvalue.} $\l_1$ of the Laplacian for any hyperbolic 2-orbifold (this includes hyperbolic surfaces) is bounded by
	\begin{align}
		\l_1<44.8883537\label{general}.
	\end{align}
	This is very close to being saturated by the $[0; 2, 3, 7]$ orbifold,\footnote{The unique genus zero orbifold with conical singularities of order 2, 3, and 7.} which has $\l_1\approx44.8883537$. In~\cite{Kravchuk:2021} it was also shown that a structure constant, analogous to an OPE coefficient in a CFT, can be bounded by
	\begin{align}
		S_{12}\gtrsim1.154096944322,\label{S12bound}
	\end{align}
	which is again almost saturated by the $[0; 2, 3, 7]$ orbifold, which has
	\begin{align}
		S_{12}^{[0; 2, 3, 7]}\approx1.154096944394.
	\end{align}
	Other bounds have also been considered in~\cite{Kravchuk:2021, Bonifacio:2021}. For example, the bound on $\l_1$ of orbifolds of genus 2 or higher is almost saturated by the Bolza surface.
	
	Similar near-saturation of bounds plays a central role in the CFT context, where it has been observed that various non-perturbative CFTs such as the 3D Ising CFT are very close to saturating bootstrap bounds~\cite{El-Showk:2012cjh, El-Showk:2014dwa}. Practically speaking, this gives rigorous and precise determinations of scaling dimensions and of other parameters of such CFTs (especially in the case of two-sided bounds). Conceptually, this offers a very tantalizing possibility that CFTs as complex as the 3D Ising CFT might exactly saturate bootstrap bounds which arise from relatively simple systems of crossing equations. Through extremal functional methods~\cite{El-Showk:2012vjm}, this would give a hope of obtaining an exact solution of such theories.
	
	In fact, some free theories are known to exactly saturate bootstrap bounds~\cite{Mazac:2016qev, Caron-Huot:2020adz}, which gives evidence in support of this possibility. Similarly, bounds on sphere-packing density~\cite{Cohn:2003} (which are equivalent to spinless modular bootstrap bounds \cite{Hartman:2019pcd}) are known to be exactly saturated in certain cases \cite{Viazovska:2017, Cohn:2017}.\footnote{And in other cases have been rigorously proven not to be saturated \cite{Cohn:2022}.} However, interacting CFTs are expected to have spectra which are much more complex than these examples, and the question of whether they can exactly saturate the bootstrap bounds — ``Does the Ising island shrink to a point?'' — remains wide open.
	
	The situation with hyperbolic orbifolds such as $[0; 2, 3, 7]$ or the Bolza surface appears to be an interesting model in this context. On the one hand, the crossing equations are closely related to those satisfied by CFTs~\cite{Kravchuk:2021} and the Laplace spectra of hyperbolic orbifolds are expected to be much more complex than those of free theories. On the other hand, investigating near-saturation of the bootstrap bounds is easier in this case, as geometric methods are available with which to calculate data about these orbifolds.
	
	In this paper, we show that under reasonable assumptions about the behaviour of numerics, the hyperbolic bootstrap bounds of the kind described above cannot be saturated by actual orbifolds. This gives an explicit example where the bootstrap bounds, although being extremely close to saturation -- in the case of the OPE coefficient $S_{12}$, to within 11 significant digits! -- eventually fall short of determining the exact spectrum.
	
	We give a brief summary of the overall argument in Section~\ref{summary}, and an outlook in Section~\ref{conclusions}. The rest of the paper is devoted to the technical details of the derivation. In Section~\ref{methods} we introduce the general methods from~\cite{Kravchuk:2021}, by which we bound the eigenvalues and OPE coefficients, and we then introduce the quantities we will use to show that the bounds are not saturated and derive methods to numerically calculate them. In Section~\ref{results}, we introduce the numerical results that show that the bootstrap bounds are indeed not saturated. We also supply an appendix in which we provide explicit metrics for $[0; k_1, k_2, k_3]$ orbifolds and for the Bolza surface.
	
	\subsection{Summary}
	\label{summary}
	
	Any hyperbolic 2-orbifold $X$ can be represented by $X=\G\backslash\H^2$, where $\H^2$ is the hyperbolic upper half-plane and $\G$ is a discrete cocompact subgroup of $\PSL(2, \R)$, which in turn is the connected component of the isometry group of $\H^2$. The group $\G$ is isomorphic to the fundamental group of $X$.
	
	The key observation~\cite{Bonifacio:2020xoc} by which we bound the eigenvalues of an orbifold, is that by considering associativity of products in $L^2(\G\backslash \PSL(2, \R))$, we can gain a series of equalities involving the spectrum of the Laplacian, which we call the crossing equations:\footnote{For details of how these equations are derived, the reader is referred to~\cite{Kravchuk:2021}.}
	\begin{align}
		\mathcal{F}^n_{2n+l}(0)+\sum_{k=1}^{\infty}T_k\mathcal{F}^n_{2n+l}(\l_k)=\begin{cases}
			S_{2n+l}&\text{if }l\text{ is even,}\\
			0&\text{otherwise},
		\end{cases}
		\label{crossing0}
	\end{align}
	for all integer $l\geq0$, where $\l_k$ are the non-zero eigenvalues of the Laplacian on the orbifold, and $S_{2n+l}$ and $T_k$ are non-negative constants called the OPE coefficients, associated with the orbifold, and $\cF^n_{2n+l}(\l)$ are a series of polynomials. We must choose $n$ depending on the class of orbifolds on which we wish to bound the spectral gap, and for the most general bound~\eqref{general}, we set $n=6$.
	
	Moreover, with a linear combination of $\L+1$ of these equations, we can get (for the general bound \eqref{general})
	\begin{align}
		\sum_{\substack{l=0\\l\text{ even}}}^{\L}(-\alpha_l) S_{2n+l}+P^n_\alpha(0)+\sum_{k=1}^{\infty}T_kP^n_\alpha(\l_k)=0,\label{crossing0.5}
	\end{align}
	where $\alpha_l$ are any coefficients we choose and $P^n_\alpha(\l)=\sum_{l=0}^\L \alpha_l\mathcal{F}^n_{2n+l}(\l)$, which we will call a functional.
	
	To find a bound, we look for a value $\tl\l_1$ and a set of coefficients $\alpha_l$ such that:
	\begin{itemize}
		\item $P^n_\alpha(\l)\geq0$ on $\big[\tl\l_1, \infty\big)$;
		\item $P^n_\a(0)>0$;
		\item for even $l$, $\alpha_l<0$.
	\end{itemize}
	If such $\alpha_l$ can be found, it follows that $\l_1\leq \tl \l_1$. Indeed, if all of the eigenvalues were in $\{0\}\cup[\tl\l_1, \infty)$, then the left-hand side of \eqref{crossing0.5} would be positive, giving us a contradiction, and so we must have $\l_1<\tl\l_1$. For a given $\L$, we then look for the smallest $\tl\l_1$ such that $\a_l$ can be found, which we shall denote by $\tl\l_1^{\L}$. We have $\l_1\leq \tl\l_1^\L$ and we cannot obtain stronger bounds at the given value of $\L$.
	
	We are interested in the asymptotic behaviour of these bounds as $\L\to\infty$. It is clear that $\tl\l_1^\L$ has a well-defined limit $\tl\l_1^\infty$ as it is decreasing and bounded below, but a priori the behaviour of $\alpha_l$ in this limit is less clear. We see numerically however, that when suitably normalized, these coefficients also seem to converge to particular values $\alpha_l^\infty$, which we call the extremal functional.
	
	Assuming the existence of an extremal functional, \eqref{crossing0.5} must hold with $\a_l=\a_l^\infty$ and $\L=\infty$. If for some given orbifold $\l_1=\tl\l_1^{\infty}$, then each of the terms in this version of \eqref{crossing0.5} is non-negative, and so they must all be zero. We can check if this is indeed the case. Firstly, we can estimate the coefficients $\a_l^\infty$ from the numerics. Secondly, we can compute the OPE coefficients such as $S_p$ for a given orbifold from geometric methods. Doing this for the general bound on $\l_1$  (in which case the candidate orbifold is $[0; 2, 3, 7]$ and $n=6$), we find that, indeed, $\a_l S_{2n+l}$ seems to be converging to 0 for $0\leq l\leq 27$. However, for $l=28$ we find that the limit $\a_{28}^\infty S_{40}$ appears to be non-zero. We can hence see that the bound cannot be saturated by this orbifold. By a more quantitative version of this argument, we show in \eqref{237nonsat} that in fact
	\begin{align}
		\tl\l_1^\infty-\l_1^{[0; 2, 3, 7]}\gtrsim1.7\times 10^{-8}.\label{237diff}
	\end{align}
	Moreover, it is known from~\cite{Kravchuk:2021} that this is the only orbifold that could saturate this bound, and so the bound cannot be saturated for any orbifold.
	
	We use similar techniques to show that the asymptotic bound on $S_{12}$ is not saturated, and we find that, denoting this bound by $\tl{S}^\infty_{12}$, 
	\begin{align}
		S_{12}-\tl{S}^\infty_{12}\gtrsim 4.1\times10^{-11},\label{OPE Non-saturation}
	\end{align}
	which is comparable to the observed difference
	\begin{align}
		S_{12}-\tl{S}^{\L=79}_{12}\approx7.2\times10^{-11}.\label{S12diff}
	\end{align}
	We also find that the asymptotic bound on $\l_1$ for genus 2 surfaces is not saturated by the Bolza surface and provide a similar bound on the difference $\l_1-\tl\l_1^\infty$.
	
	In order to quantify the non-saturation of these bounds, we derive methods to calculate OPE coefficients on specific orbifolds. For the coefficients $T_k$ we use the numerical bootstrap to find two-sided bounds on them. For the coefficients $S_p$, we relate them to the Petersson norms of Rankin-Cohen brackets of modular forms. These norms are calculated explicitly by finding models of the relevant orbifolds in terms of punctured spheres with explicit metrics, which we do in the cases of the $[0; 2, 3, 7]$ orbifold and the Bolza surface.
	
	\section{Numerical Bootstrap\label{methods} and OPE Coefficients}
	\subsection{Setup and Bounds\label{setup}}
	We consider the hyperbolic orbifold $X=\G\backslash\H^2$ for some discrete cocompact subgroup $\G<G\equiv\PSL(2, \R)$. We shall be interested in the spectrum of the Laplacian, or explicitly in solutions to 
	\begin{align}
		-\Delta f\equiv-\sum_{ij}\frac{1}{\sqrt{\det g}}\partial_i(\sqrt{\det g}\, g^{ij}\partial_j f)=\l f,
	\end{align}
	where $g$ is the unique Riemannian metric of constant Gaussian curvature $-1$ on $X$ (henceforth referred to as the hyperbolic metric).
	
	The eigenvalues $\l$ can be placed in non-decreasing order as
	\begin{align}
		0=\l_0<\l_1\leq\l_2\leq\dots
	\end{align}
	and we are interested in finding bounds $\tl\l_1$ such that for any choice of $\G$,
	\begin{align}
		\l_1\leq \tl\l_1.
	\end{align}
	
	We define the signature of $X$ to be $[g; k_1, k_2,\dots, k_r]$ if $X$ is of genus $g$ and has $r$ orbifold points of order $k_1\dots k_r$. Let $\ell_n$ be the dimension of space of holomorphic sections of the $n$\textsuperscript{th} power of the canonical bundle over $X$, to which we will refer as the holomorphic differentials of degree $n$. The Riemann-Roch formula (\cite{Milne:1997}, Theorem 4.9) tells us that 
	\begin{align}
		\ell_n=(2n-1)(g-1)+ \sum_{i=1}^r \left\lfloor n\frac{k_i-1}{k_i}\right\rfloor +\delta_{n,1}\label{RR}.
	\end{align}
	
	\paragraph{Single-correlator bounds} For any $n$ such that $\ell_n\geq1$, we have at least one holomorphic differential of degree $n$. A crossing equation for this holomorphic differential can be formulated following~\cite{Kravchuk:2021,Bonifacio:2021}, which gives us the constraint that for any $l\geq0$,
	\begin{align}
		\mathcal{F}^n_{2n+l}(0)+\sum_{k=1}^{\infty}T_k\mathcal{F}^n_{2n+l}(\l_k)=\begin{cases}
			S_{2n+l}&\text{if }l\text{ is even,}\\
			0&\text{otherwise},
		\end{cases}
		\label{crossing00}
	\end{align}
	where
	\begin{align}
		\cF^n_p(\l)=\sum_{a+b+c=p-2n}\frac{(-1)^a(2n+a)_c(1-p)_b^2}{c!(2-2p)_b b! (a!)^2}\prod_{k=0}^{a-1}(\l+k+k^2).\label{polys}
	\end{align}
	The sum runs over all non-negative integers $a, b, c$ that sum to $p-2n$, and $S_{2n+l}$, and $T_k$ are non-negative constants. The $T_k$ are proportional to the squared overlap integral of the holomorphic differential, its complex conjugate, and a Laplace eigenfunction~\cite{Kravchuk:2021}. The $S_{2n+l}$ are proportional, as we show in Section~\ref{definitons}, to the squared structure constants of the Rankin-Cohen algebra~\cite{Zagier:1994} of holomorphic differentials on $X$.
	
	Using the Riemann-Roch formula \eqref{RR} it can be shown that one of $\{\ell_1, \ell_2, \ell_3, \ell_4, \ell_6\}$ is always non-zero~\cite{Kravchuk:2021}. By considering the bounds arising from each of these cases separately, we find that $n=6$ gives the weakest and thus the most general bound \eqref{general}.
	
	Considering a linear combination of the conditions \eqref{crossing00} with coefficients $\a_l$, we obtain \eqref{crossing0.5}. For a given, $\tl\l_1$, we search for a set of coefficients $\a_l$ such that setting $P^n_\alpha(\l)=\sum_{l=0}^\L \alpha_l\mathcal{F}^n_{2n+l}(\l)$, we have
	\begin{align}\label{eq:alphaconditions}
		P^n_\alpha(0)=1;&\nn\\
		P^n_\alpha(\l)\geq0,& \qquad \text{for all }\l\geq\tl\l_1;\\
		\alpha_l\leq0,& \qquad \text{for $l$ even};\nn
	\end{align}
	as this gives a contradiction if $\l_1\geq \tl\l_1$, following the standard arguments outlined in the introduction.
	
	This search can be performed numerically using the semidefinite program solver \verb+SDPB+~\cite{Simmons-Duffin:2015}, and the optimal bound $\tl\l_1$ can by found by performing a binary search. Applying this to $n=6$ with $\Lambda=77$, we find that
	\begin{align}\label{eq:best237gap}
		\tl\l_1^{\Lambda=77}\approx44.8883536337,
	\end{align}
	which agrees with the $\Lambda=41$ bound from~\cite{Kravchuk:2021} stated in \eqref{general}.
	
	\paragraph{Multi-correlator bounds} For any genus $g$ orbifold, \eqref{RR} gives that $\ell_1=g$. This gives us $g$ holomorphic differentials of degree 1 with which we can form a system of crossing equations which give us the constraint~\cite{Kravchuk:2021} that for $p\geq 2n$,
	\begin{align}
		S_p=\mathcal{F}^1_p(0)+\sum_{k=1}^{\infty}\p{Q_k-\frac{1}{\ell_1}T_k}\mathcal{F}^1_p(\l_k)=\sum_{k=1}^{\infty}T_k\mathcal{F}^1_p(\l_k)\quad&\text{if }p\text{ is even};\\
		A_p=\mathcal{F}^1_p(0)+\sum_{k=1}^{\infty}\p{Q_k-\frac{1}{\ell_1}T_k}\mathcal{F}^1_p(\l_k)=-\sum_{k=1}^{\infty}T_k\mathcal{F}^1_p(\l_k)\quad&\text{if }p\text{ is odd};
		\label{crossing2ab}
	\end{align}
	where $S_p, A_p, T_k,$ and $Q_k$ are non-negative coefficients, which we shall once again call the OPE coefficients. The $T_k$ and $Q_k$ are proportional to the squared overlap integral of the holomorphic differentials, their complex conjugates, and a Laplace eigenfunction, which have been symmetrized and anti-symmetrized respectively. $S_p$ and $A_p$ are similarly proportional to squared symmetrized and anti-symmetrized structure constants of the Rankin-Cohen algebra.
	
	As before, we consider a linear combination of these conditions,
	\begin{align}
		&-\sum_{\substack{l=0\\l\text{ even}}}^{\Lambda}\alpha_lS_{2+l}-\sum_{\substack{m=1\\l\text{ odd}}}^{\Lambda}\beta_m A_{2+m}+P^1_{\gamma}(0)+P^1_{\delta}(0)+\sum_{k=1}^{\infty}Q_k(P^1_{\gamma}(\l_k)+P^1_{\delta}(\l_k))\nn\\
		&+\sum_{k=1}^{\infty}\left(T_k\left(P^1_{\alpha}(\l_k)-P^1_{\beta}(\l_k)-\left(\frac{\ell_1+1}{\ell_1}\right)P^1_{\gamma}(\l_k)+\left(\frac{\ell_1-1}{\ell_1}\right)P^1_{\delta}(\l_k)\right)\right)=0,\label{multi}
	\end{align}
	where $\alpha_l$ and $\gamma_l$ are only non-zero for even $l$, and $\beta_m$ and $\delta_m$ are only non-zero for odd $m$.
	
	Given a potential $\tl\l_1$, we once again use a combination of binary search and semidefinite programming to look for values of these coefficients such that every term in the crossing equations is positive. Applying this process for $g=2$, $\L=59$, we obtain a bound that is almost saturated by the Bolza surface,
	\begin{align}
		\tl\l_1^{\Lambda=59, \, g=2}\approx 3.8388976481487.\label{BolzaBound}
	\end{align}
	
	\paragraph{OPE bounds} We can use a similar approach to bound a particular OPE coefficient, by isolating the term containing it in the crossing equation. For example, in the case of the single-correlator bound from~\eqref{crossing00} we can bound $S_{2n}$, by writing~\eqref{crossing0.5} as
	\begin{align}
		\alpha_0 S_{2n}-P^n_\alpha(0)=\sum_{\substack{l=2\\l\text{ even}}}^{\L}(-\alpha_l) S_{2n+l}+\sum_{k=1}^{\infty}T_kP^n_\alpha(\l_k).\label{OPE Crossing}
	\end{align}
	
	If we find a functional such that
	\be\label{eq:conditions2}
	\alpha_0=1;&\nn\\
	\alpha_l\leq0,&\qquad\text{for even $l\geq2$};\\
	P^n_\a(\l)\geq0,&\qquad\text{ for $\l\geq0$};\nn
	\ee
	that maximizes $P^n_\a(0)$, then we find that $S_{2n}\geq P^n_\a(0)$. A similar argument can be used to construct bounds for $T_k$.
	
	Denoting this lower bound by $\tl{S}_{2n}^{\Lambda}$ and applying this with $n=6$, $\Lambda=79$ gives us
	\begin{align}
		\tl{S}_{12}^{\Lambda=79}\approx1.1540969443224
	\end{align}
	which agrees with the $\Lambda=37$ bound from~\cite{Kravchuk:2021} stated in \eqref{S12bound}.
	
	We can also use this to bound eigenvalues and OPE coefficients on concrete orbifolds by noting that we only actually need $P^n_\a(\l)$ to be non-negative at each $\l_k$. Hence, if we have some information about the spectrum, then we can often obtain tight two-sided bounds on these quantities. Additionally, we may be able to use the Riemann-Roch formula to show that certain OPE coefficients are zero, allowing us to relax the negativity constraints on the corresponding $A_p$ and $S_p$. We will later use this idea to effectively compute $T_1$ on concrete orbifolds, which is necessary to quantify the non-saturation of the bootstrap bounds, \eqref{general}, \eqref{S12bound} and \eqref{BolzaBound}.
	
	\subsection{Definitions of the OPE coefficients}
	\label{definitons}
	
	In this section we give the precise definition of the OPE coefficients $A_p$ and $S_p$ which will be required in what follows. This section is essentially a review of the relevant parts of~\cite{Kravchuk:2021}.
	
	With $G=\PSL(2, \R)$, let $\G<G$ be the subgroup such that $\G\backslash \H^2$ is isometric to the 2-orbifold $X$. We recall that we can use the Iwasawa decomposition to parameterize elements of $G$ as
	\be
	g(x, y, \th)=\pm \begin{pmatrix}1&x\\0&1\end{pmatrix}\begin{pmatrix}\sqrt{y}&0\\0&\frac{1}{\sqrt{y}}\end{pmatrix}\begin{pmatrix}\cos\frac{\th}{2}&-\sin\frac{\th}{2}\\\sin\frac{\th}{2}&\cos\frac{\th}{2}\end{pmatrix},
	\ee
	where $x\in(
	-\infty, \infty), y\in(0, \infty), \th\in\R/2\pi\Z$, and $\pm$ is needed because we are working with $\PSL(2,\R)$ and not $\SL(2,\R)$.
	
	The Haar measure on $G$ is given by
	\begin{align}
		d\mu(g(x, y, \th))=\frac{1}{2\pi\vol(\Gamma\backslash\H^2)}\frac{dx\,dy\,d\th}{y^2}.
	\end{align}
	This measure is both left-invariant and right-invariant, and we shall normalize it by setting $\mu(\Gamma\backslash G)=1$. $L^2(\Gamma\backslash G)$ is then defined to be the space of functions $F: \Gamma\backslash G\to\C$ such that $\int_{\Gamma\backslash G}|F(g)|^2\,d\mu(g)<\infty$. This is a Hilbert space with the inner product given by:\footnote{We follow the standard convention from the physics literature, and define this inner product to be linear in the second argument.} $\p{F_1, F_2}=\int_{\Gamma\backslash G}F_1(g)^*\,F_2(g)\,d\mu(g)$, and an induced norm given by $\|F\|=\sqrt{\p{F, F}}$. We shall often lift $F$ to a function from $G$ to $\C$, and by abuse of notation, we shall also call this function $F$.
	
	We can introduce a basis for the complexified Lie algebra of $G$, $\mathfrak{g}_{\C}$, which acts on $L^2(\Gamma\backslash G)$, as
	\be
	(L_{-1}F)(x, y, \theta)&=e^{-i\theta}[y(\partial_x-i\partial_y)+\partial_{\theta}]F(x, y, \theta),\nn\\
	(L_0F)(x, y, \theta)&=i\partial_{\theta}F(x, y, \theta),\\
	(L_1F)(x, y, \theta)&=-e^{i\theta}[y(\partial_x+i\partial_y)+\partial_{\theta}]F(x, y, \theta),\nn
	\ee
	and which obeys the commutation relations $[L_m, L_n]=(m-n)L_{m+n}$.
	
	We can decompose $L^2(\Gamma\backslash G)$ into eigenspaces of $L_0$,
	\be
	L^2(\Gamma\backslash G)=\bigoplus_{n\in\Z}V_n,
	\ee
	where $L_0=n$ on $V_n$. We parameterize elements of $V_n$ as
	\be
	F(x,y,\theta)=y^{|n|}e^{-in\th}f_n(x,y).
	\ee
	
	Moreover, if we switch to complex coordinates, $z=x+iy$, $\bar{z}=x-iy$, and let $h_n(z, \bar{z})=f_n(x, y)$, we can see that invariance of the functions $F$ under $\G$ requires that
	\be
	h_n\p{\g\cdot z,\bar{\g \cdot z}} =
	\begin{cases}
		(c+dz)^{2n}h_n(z, \bar{z})&\text{if }n\geq0\\
		(c+d\bar{z})^{-2n}h_n(z, \bar{z})&\text{if }n<0
	\end{cases},\quad\quad\forall\, \gamma=\begin{pmatrix}a&b\\c&d\end{pmatrix}\in\G,
	\ee
	where $\g\cdot z=\frac{az+b}{cz+d}$.
	
	We define for $n\geq0$
	\begin{align}
		W_n=\{y^{|n|}e^{-in\th}h_n(x+iy)\in V_n\;|\;h_n\text{ holomorphic}\}.
	\end{align}
	The above discussion implies that $h_n$ is a holomorphic modular form of weight $2n$ of level $\G$, and so $W_n$ is isomorphic to the space of holomorphic modular forms of weight $2n$ of level $\G$, $M_{2n}(\Gamma)$. Explicitly calculating the inner product of these functions in $L^2(\G\backslash G)$ for $n\geq0$, we find that,
	\begin{align}
		&\langle y^{n}e^{-in\th} g_n(x+iy), y^{n}e^{-in\th} h_n(x+iy)\rangle_{L^2(\G\backslash G)}\nn\\
		&=\frac{1}{\vol(\Gamma\backslash\H^2)}\int_{\G\backslash\H^2}  \bar{g_n(x+iy)}\,h_n(x+iy)y^{2n-2}\,dy\,dx\label{Petersson}\\
		&=\langle g_n, h_n\rangle_{M_{2n}(\Gamma)},\nn
	\end{align}
	which is the Petersson inner product of modular forms. We shall denote its associated norm by $\|h_n\|_{M_{2n}(\Gamma)}=\sqrt{\langle h_n, h_n\rangle_{M_{2n}(\Gamma)}}$.
	
	As the differentials $h_n(z, \bar{z})dz^n$ are invariant under the action of $\G$, they constitute the space of holomorphic differentials of degree $n$ on $X$, and so $\ell_n$ is the dimension of $M_{2n}(\G)$. We shall henceforth assume that $\{h_{n, a}(z)\}_{a=1}^{\ell_n}$ is an orthonormal basis for $M_{2n}(\G)$ with respect to the Petersson inner product \eqref{Petersson}.
	
	Decomposing $L^2(\G\backslash G)$ into irreducible representations of $G$, we find that the functions in $W_n$ correspond to lowest-weight vectors in discrete series representations. By analogy with the representation theory of the conformal algebra, we can then define what we shall call coherent states, which are analogues of primary operators in a CFT. Given one of the basis elements $h_{n, a}$, we define the associated holomorphic discrete series coherent state by
	\begin{align}
		\mathscr{O}_{n, a}(w)(x, y, \theta)=\exp(wL_{-1})\left(y^ne^{-in\th}h_{n, a}(x+iy)\right),
	\end{align}
	where $w\in \mathbb{C}, |w|<1$.
	
	We also define an anti-holomorphic coherent state\footnote{This state is still meromorphic in $w$, but we call it an anti-holomorphic coherent state as it depends on the anti-holomorphic modular form $\bar{h_{n, a}(x+iy)}$.} by
	\begin{align}
		\tilde{\mathscr{O}}_{n, a}(w)(x, y, \theta)=w^{-2n}\exp(-w^{-1}L_{-1})\left(y^ne^{in\th}\bar{h_{n, a}(x+iy)}\right),
	\end{align}
	where $w\in \mathbb{C}, |w|>1$.
	
	Extending the analogy with CFTs, we can define correlators of $k$ functions $F_i\in \Gamma\backslash G$ as
	\begin{align}
		\left\langle F_1 F_2 \dots F_k \right\rangle=\int d\mu(g) F_1(g) F_2(g) \dots F_k(g),
	\end{align}
	and similarly to CFT two-point functions, it can be seen that
	\begin{align}
		\left\langle \mathscr{O}_{n_1, a_1}(w_1) \tl{\mathscr{O}}_{n_2, a_2}(w_2) \right\rangle=\frac{\delta_{n_1, n_2} \delta_{a_1, a_2}}{(w_1-w_2)^2}.
	\end{align}
	
	By projecting a product of these coherent states $\mathscr{O}_{n, a_1}(w_1)\mathscr{O}_{n, a_2}(w_2)$ onto the irreducible representations of $G$ in $L^2(\G\backslash G)$, we can introduce the so-called operator product expansion (OPE) between these coherent states, which in~\cite{Kravchuk:2021} was found to be
	\begin{align}
		\mathscr{O}_{n, a_1}(w_1)\,\mathscr{O}_{n, a_2}(w_2)=\sum_{p=2n}^{\infty}\sum_{a=1}^{\ell_p}\sum_{m=0}^{\infty}\frac{(p)_{m}}{(2p)_{m}m!}f^{a_1, a_2}_{p, a}(w_1-w_2)^{p+m-2n}L_{-1}^m\mathscr{O}_{p, a}(w_2),\label{OPE}
	\end{align}
	where $f^{a_1, a_2}_{p, a}$ are coefficients depending on the choice of $\Gamma$, and $(p)_m$ denotes the Pochhammer symbol $(p)_m=\prod_{k=0}^{m-1}(p+k)$.
	
	The crossing equations \eqref{crossing0} and \eqref{multi} are derived by considering the correlator 
	\be
	\langle \mathscr{O}_{n, a_1}(w_1)\,\mathscr{O}_{n, a_2}(w_2)\,\tl{\mathscr{O}}_{n, a_3}(w_3)\,\tl{\mathscr{O}}_{n, a_4}(w_4)\rangle\label{corr},
	\ee
	expanding it into a sum of ``conformal blocks'' using OPEs between different pairs of fields, and imposing equality between these expansions.\footnote{Specifically, we use an OPE between $\mathscr{O}_{n, a_1}(w_1)\,\mathscr{O}_{n, a_2}(w_2)$ and between $\tl{\mathscr{O}}_{n, a_3}(w_3)\,\tl{\mathscr{O}}_{n, a_4}(w_4)$ on one side of the equation, and on the other side between $\mathscr{O}_{n, a_1}(w_1)\,\tl{\mathscr{O}}_{n, a_3}(w_3)$ and between $\mathscr{O}_{n, a_2}(w_2)\,\tl{\mathscr{O}}_{n, a_4}(w_4)$. This requires expressions for other OPEs which are given in~\cite{Kravchuk:2021}.}
	
	Normally, the crossing equations would involve the individual OPE coefficients such as $f_{p,a}^{a_1,a_2}$. However, since for a fixed $n$ all the $\ell_n$ coherent states $\mathscr{O}_{n,a}$ transform in the same representation, there is an enhancement of symmetry in the bootstrap bounds. This allows a reduction (equivalent from the point of view of the bootstrap bounds) of the full system of crossing equations to the simple systems \eqref{crossing0} and \eqref{multi} we described in Section~\ref{setup}.
	
	To obtain \eqref{crossing0}, we consider the case where $a_1=a_2=a_3=a_4$ in \eqref{corr}, and the OPE coefficients $S_p$ are given by
	\begin{align}
		S_{p}=\sum_{a=1}^{\ell_p}|f^{a_1, a_1}_{p, a}|^2.\label{Sp1}
	\end{align}
	
	For \eqref{multi}, the OPE coefficients $A_p$ and $S_p$ are related to $f_{p,a}^{a_1,a_2}$ through
	\begin{align}
		S_p=\sum_{a=1}^{\ell_p}\sum_{a_1=1}^{\ell_n}\sum_{a_2=1}^{\ell_n}\frac{1}{\ell_n(\ell_n+1)}|f^{(a_1, a_2)}_{p, a}|^2,\label{Sp2}
	\end{align}
	and
	\begin{align}
		A_p=\sum_{a=1}^{\ell_p}\sum_{a_1=1}^{\ell_n}\sum_{a_2=1}^{\ell_n}\frac{1}{\ell_n(\ell_n-1)}|f^{[a_1, a_2]}_{p, a}|^2.\label{Ap2}
	\end{align}
	Here, the symmetric and anti-symmetric parts of $f^{a_1, a_1}_{p, a}$ are defined as
	\begin{align}
		f^{(a_1, a_2)}_{p, a}&=\frac{1}{2}\p{f^{a_1, a_2}_{p, a}+f^{a_2, a_1}_{p, a}},\\
		f^{[a_1, a_2]}_{p, a}&=\frac{1}{2}\p{f^{a_1, a_2}_{p, a}-f^{a_2, a_1}_{p, a}}.
	\end{align}
	
	\subsection{OPE coefficients from Rankin-Cohen brackets}
	
	In this section we relate the OPE coefficients $S_p$ and $A_p$ to the structure constants of a Rankin-Cohen algebra. In Section~\ref{OPE Calculations}, we shall then show how this can be used to calculate these OPE coefficients on specific orbifolds, such as the $[0; 2, 3, 7]$ orbifold and the Bolza surface, using the metrics given in Appendix~\ref{appendix}.
	
	We shall introduce the notation
	\begin{align}
		\hat{\mathscr{O}}^{a_1, a_2}_{p}(w)=\sum_{a=1}^{\ell_p}f^{a_1, a_2}_{p, a}\mathscr{O}_{p, a}(w)
	\end{align}
	and
	\begin{align}
		\hat{h}_p^{a_1, a_2}(z)=\sum_{a=1}^{\ell_p}f^{a_1, a_2}_{p, a}h_{p, a}(z)
	\end{align}
	in the discussion that follows. This notation is helpful, as by orthonormality of our basis of modular forms with respect to \eqref{Petersson}, we can write the OPE coefficient \eqref{Sp1} that appears in our single correlator crossing equation \eqref{crossing0} as
	\begin{align}
		S_p=\left\|\hat{h}^{a_1, a_1}_{p}\right\|_{M_{2n}(\Gamma)}^2=\left\|\hat{\mathscr{O}}^{a_1, a_1}_{p}(w)\right\|_{L^2(\Gamma\backslash G)}^2.\label{SpNormForm}
	\end{align}
	
	Similarly for the multi-correlator bound \eqref{multi}, we can write the OPE coefficients as
	\begin{align}
		S_p=\sum_{a_1=1}^{\ell_n}\sum_{a_2=1}^{\ell_n}\frac{1}{\ell_n(\ell_n+1)}\left\|\hat{h}^{(a_1, a_2)}_{p}\right\|_{M_{2n}(\Gamma)}^2=\sum_{a_1=1}^{\ell_n}\sum_{a_2=1}^{\ell_n}\frac{1}{\ell_n(\ell_n+1)}\left\|\hat{\mathscr{O}}^{(a_1, a_2)}_{p}(w)\right\|_{L^2(\Gamma\backslash G)}^2,
	\end{align}
	and
	\begin{align}
		A_p=\sum_{a_1=1}^{\ell_n}\sum_{a_2=1}^{\ell_n}\frac{1}{\ell_n(\ell_n-1)}\left\|\hat{h}^{[a_1, a_2]}_{p}\right\|_{M_{2n}(\Gamma)}^2=\sum_{a_1=1}^{\ell_n}\sum_{a_2=1}^{\ell_n}\frac{1}{\ell_n(\ell_n-1)}\left\|\hat{\mathscr{O}}^{[a_1, a_2]}_{p}(w)\right\|_{L^2(\Gamma\backslash G)}^2\label{ApNormForm}
	\end{align}
	In order to calculate the OPE coefficients, we shall hence look for formulas for $\hat{\mathscr{O}}^{a_1, a_2}_{p}(w)$ and $\hat{h}^{a_1, a_2}_{p}$.
	
	The starting point for our explicit formulas, is to rewrite \eqref{OPE} in terms of $\hat{\mathscr{O}_p}^{a_1, a_2}$,
	\begin{align}
		\mathscr{O}_{n, a_1}(w_1)\,\mathscr{O}_{n, a_2}(w_2)=\sum_{p=2n}^{\infty}\sum_{m=0}^{\infty}\frac{(p)_{m}}{(2p)_{m}m!}(w_1-w_2)^{p+m-2n}L_{-1}^m\hat{\mathscr{O}}_p^{a_1, a_2}(w_2).
	\end{align}
	Expanding this equation around the point $w_1=w_2$, we can invert it to obtain expressions for the coherent states $\hat{\mathscr{O}}^{a_1, a_2}_{p}(w)$.  For example, by expanding to second order in $w_1-w_2$, we can find that
	\begin{align}
		\hat{\mathscr{O}}^{a_1, a_2}_{2n}(w_2)&=\mathscr{O}_{n, a_1}(w_2)\,\mathscr{O}_{n, a_2}(w_2),\\
		\hat{\mathscr{O}}^{a_1, a_2}_{2n+1}(w_2)&= \frac{\partial}{\partial w_1}\mathscr{O}_{n, a_1}(w_1)\,\mathscr{O}_{n, a_2}(w_2) \,\Bigg|_{w_1=w_2}-\frac{1}{2} L_{-1} \hat{\mathscr{O}}^{a_1, a_2}_{2n}(w_2)\\
		\hat{\mathscr{O}}^{a_1, a_2}_{2n+2}(w_2)&=\frac{1}{2}\frac{\partial^2}{\partial w_1^2}\mathscr{O}_{n, a_1}(w_1)\,\mathscr{O}_{n, a_2}(w_2) \,\Bigg|_{w_1=w_2}-\frac{1}{2}L_{-1}\hat{\mathscr{O}}^{a_1, a_2}_{2n+1}(w_2)\\
		&\qquad-\frac{(2n+1)}{4 (4 n+1)}L_{-1}^2 \hat{\mathscr{O}}^{a_1, a_2}_{2n}(w_2)\nn.
	\end{align}
	This is analogous to the way we can create new primary operators from an OPE in mean field theory.
	
	Considering the way that these identities look in terms of $h_{n, a}$, we in fact find that we have a bilinear differential operator of degree $2k$ from $M_{2n}(\Gamma)\times M_{2n}(\Gamma)$ to $M_{4n+2k}(\Gamma)$. The unique such map (up to a multiplicative constant) is the Rankin-Cohen bracket~\cite{Zagier:1994, Kravchuk:2021}, which is defined by
	\be\label{eq:RCdefn}
	[h_1, h_2]_k(z) =\sum_{r+s=k}(-1)^r\binom{2n+k-1}{s}\binom{2n+k-1}{r}h_1^{(r)}(z)h_2^{(s)}(z).
	\ee
	In fact, we can prove by induction that the two constructions are related by
	\begin{align}
		\hat{h}_{2n+k}^{a_1, a_2}(z)=\frac{2^k}{(4n+k-1)_{k}}[h_{n, a_1}, h_{n, a_2}]_k(z).\label{newprims}
	\end{align}
	
	Noting that the Rankin-Cohen bracket is symmetric for $k$ even, and anti-symmetric for $k$ odd, we can find that $\hat{h}_{2n+k}^{[a_1, a_2]}(z)=0$ if $k$ is even, and $\hat{h}_{2n+k}^{(a_1, a_2)}(z)=0$ if $k$ is odd. We will show an explicit way to compute $\hat{h}_{p}^{a_1, a_2}$ for the $[0; 2, 3, 7]$ orbifold and the Bolza surface in the next section and in Appendix~\ref{appendix}.
	
	\subsection{Rankin-Cohen Algebra on Punctured Spheres\label{OPE Calculations}}
	
	We now have explicit formulas for the OPE coefficients in terms of modular forms, but unfortunately for a generic $\Gamma$, constructing such modular forms is difficult. We shall hence find an alternate framework for calculating the OPE coefficients.
	
	Noting that modular forms are just holomorphic differentials on $X=\Gamma\backslash\mathbb{H}^2$, we can look for an alternate description of $X$, where these holomorphic differentials can be more easily constructed. We shall call this alternative description $\tl{X}$. For any $[0; k_1, k_2, k_3]$ orbifold, we can take $\tl{X}$ to be a three-punctured sphere with the hyperbolic metric given in \eqref{metric}. For the $[0;2,3,7]$ orbifold this is applicable directly. The Bolza surface is the eightfold cover of the $[0; 4; 4; 4]$ orbifold, and so we can use the covering map to lift \eqref{metric} to a metric on the Bolza surface. This construction of $\tl{X}$ for the Bolza surface is detailed in Appendix~\ref{Bolza}.
	
	The holomorphic differentials will have simple expressions in local coordinates on $\tl X$ (as we shall discuss in Appendix~\ref{appendix}), and this approach gives us explicit expressions for the hyperbolic metric. The last ingredient that we require is the expression for the Rankin-Cohen bracket, which in local coordinates on $\tl X$ will be different from~\eqref{eq:RCdefn}. To derive it we will use the biholomorphic isometry between $\tl{X}$ and $X$, which we can lift to a multi-valued function $\tau: \tl{X}\to \H^2$.
	
	A modular form of weight $2n$, $h_{n, a}$ gives us a holomorphic differential $h_{n, a}(\tau)d\tau^n$ on $X$. We can then use $\tau$ to write the corresponding differential on $\tl{X}$, by
	\begin{align}
		\tl{h}_{n, a}(z)dz^n=h_{n, a}(\tau(z))d\tau(z)^n.\label{convert}
	\end{align}
	
	We shall introduce the notation $R_n$ for the space of holomorphic differentials of degree $n$ on $\tilde{X}$. For $\tl{f}(z)dz^m\in R_m$, $\tl{g}(z)dz^n\in R_n$, the differential corresponding to the Rankin-Cohen bracket is given by
	\begin{align}
		[\tl{f}(z)dz^m, \tl{g}(z)dz^n]_{D, k}(z)=\!\!\sum_{r+s=k}\!\!(-1)^r\binom{2m+k-1}{s}\binom{2n+k-1}{r}(D^r\tl{f}(z)dz^m)(D^s\tl{g}(z)dz^n),\label{RCXhat}
	\end{align}
	where
	\begin{align}
		D \tl{f}(z)dz^m=f'(\tau(z))\tau'(z)d\tau(z)^{m+1}.
	\end{align}
	As we will see later, the holomorphic differentials on $\tl X$ become simple rational functions in $z$ coordinate. In particular, $\tl f(z), \tl g(z)$, as well as the sum in the right-hand side of~\eqref{RCXhat} have simple rational expressions. On the other hand, the map $\tau$ that enters into this sum has a much more complicated expression, similar in flavour to the metric~\eqref{metric}. This means that there are some subtle cancellations that must happen in~\eqref{RCXhat}.
	
	This is manifest in the following equivalent form of \eqref{RCXhat},\footnote{This follows from Proposition 2 of~\cite{Zagier:1994}, with $F=dz$.}
	\begin{align}
		[\tl{f}(z)dz^m, \tl{g}(z)dz^n]_{D, k}(z)=\sum_{r+s=k}(-1)^r\binom{2m+k-1}{s}\binom{2n+k-1}{r}f_r(z) g_s(z)dz^{m+n+k},\label{RCDeformed}
	\end{align}
	where $f_r$ and $g_r$ are defined inductively by $f_0(z)=\tilde{f}(z)$, $g_0(z)=\tilde{g}(z)$ and
	\begin{align}
		f_{r+1}(z)&=f_r'(z)+\frac{r(1-2m-r)}{2} \{\tau(z), z\} f_{r-1}(z),\\
		g_{s+1}(z)&=g_s'(z)+\frac{s(1-2m-s)}{2} \{\tau(z), z\} g_{s-1}(z).
	\end{align}
	
	The only dependence on $\tau$ in this formula is in the Schwarzian derivative of $\tau$,
	\begin{align}
		\{\tau(z), z\}=\frac{\tau'''(z)}{\tau'(z)}-\frac{3}{2}\left(\frac{\tau''(z)}{\tau'(z)}\right)^2.\label{Schwarzian}
	\end{align}
	While $\tau(z)$ is relatively complicated, it is well-known that $\{\tau(z), z\}$ is a single-valued\footnote{The single-valuedness of this follows from the general property that the Schwarzian derivative is invariant under M\"obius transformations.} holomorphic function of $z$, with the only singularities being poles at the orbifold points. Furthermore, the leading coefficients at these poles can be expressed in terms of the angle deficits. Using this, the whole function $\{\tau(z), z\}$ can be fully determined in the case of three singularities. We give explicit formulas for it in the cases of the $[0; 2, 3, 7]$ orbifold and the Bolza surface in Appendix~\ref{appendix}.
	
	We now finally have all the necessary ingredients to compute the OPE coefficients $A_p$ and $S_p$ on the $[0; 2, 3, 7]$ orbifold and the Bolza surface. We give the remaining technical details in Appendix~\ref{appendix}.
	
	\section{Results\label{results}}

	In this section we shall finally show that the nearly-saturated bounds, \eqref{general}, \eqref{S12bound} and \eqref{BolzaBound} are not in fact saturated.
	
	\subsection{Non-saturation of $\tl{S}^\infty_{12}$ by the $[0; 2, 3, 7]$ Orbifold}
	\begin{figure}[t]
		\caption{}
		\begin{subfigure}[t]{0.43\textwidth}
			\includegraphics[width=\linewidth]{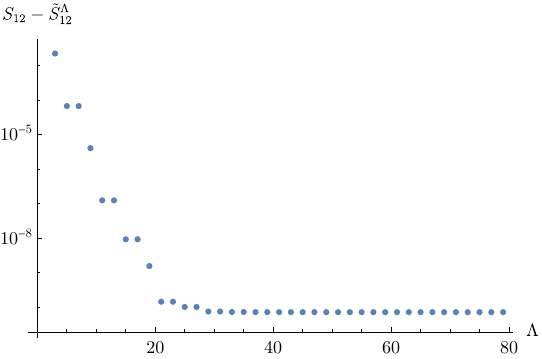}
			\caption{Plot of the difference between $S_{12}$ on the $[0; 2, 3, 7]$ orbifold, and the bounds $\tl{S}^\Lambda_{12}$.}
			\label{fig:S12Bounds}
		\end{subfigure}
		\hfill
		\begin{subfigure}[t]{0.52\textwidth}
			\includegraphics[width=\linewidth]{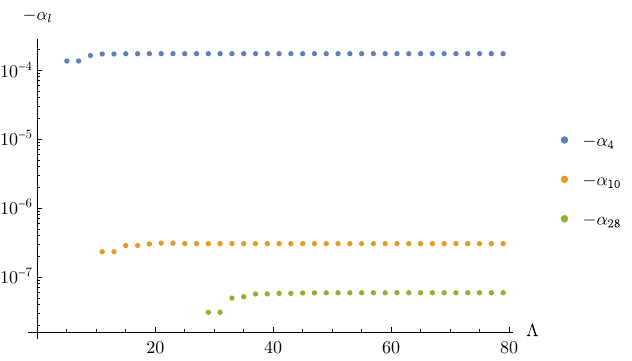}
			\caption{Plot showing the convergence of $-\alpha_4$, $-\alpha_{10}$, and $-\alpha_{28}$ to non-zero values when bounding $S_{12}$.}
			\label{fig:S12Alphas}
		\end{subfigure}
	\end{figure}
	As we noted in the introduction, we know that for $n=6$, the best lower bound we obtain on $S_{12}$ is
	\begin{align}
		\tl{S}_{12}^{\Lambda=79}\approx1.1540969443224,
	\end{align}
	which is almost saturated by the $[0; 2, 3, 7]$ orbifold, which we can calculate by \eqref{Sp},
	\begin{align}
		S_{12}\approx1.1540969443945.
	\end{align}
	We are hence interested in the question of whether the bound $\tl{S}_{12}^\infty=\lim_{\Lambda\to\infty}\tl{S}_{12}^{\Lambda}$ is exactly saturated by this orbifold. The first piece of evidence that the bound may not be saturated can be seen from Fig. \ref{fig:S12Bounds}, where we can see that while the difference $S_{12}-\tl{S}^\Lambda_{12}$ is very small, it appears to be converging to a non-zero value. We shall however assume that the bound is indeed saturated, seeking a contradiction.
	
	We note that in this limit the coefficients $\alpha_l$ appear to be converging to an extremal functional $\alpha^\infty_l=\lim_{\Lambda\to\infty}\alpha^\Lambda_l$. If the bound $\tl{S}_{12}^\infty$ is saturated by the value of $S_{12}$ for $[0;2 ,3, 7]$, then every term $\alpha_l^\infty S_{2n+l}$ in \eqref{crossing0.5} evaluated on $\a_l^\infty$ must equal zero (other than $\alpha_0^\infty S_{12}$), as they must be non-negative because of \eqref{eq:conditions2}.
	
	We illustrate the convergence of the coefficients $\alpha_l$ in Figure~\ref{fig:S12Alphas} for $\alpha_4$, $\alpha_{10}$, and $\alpha_{28}$. From the figure we can see that these coefficients appear to have essentially converged to their limiting values $\alpha_l^\infty$, and furthermore that the limiting values $\alpha^\infty_4$, $\alpha^\infty_{10}$ and $\alpha^\infty_{28}$ are non-zero. We will assume from now on that the extremal functional $\alpha^\infty_l$ for this bound exists, is unique, and that our numerics correctly approximate its values.
	
	For the bound to be saturated, the non-vanishing of $\alpha^\infty_4$, $\alpha^\infty_{10}$ and $\alpha^\infty_{28}$ implies that the corresponding OPE coefficients $S_{16},\, S_{22}$ and  $S_{40}$ must vanish on $[0; 2, 3, 7]$. The vanishing of $S_{16}$ and $S_{22}$ can be seen simply from the Riemann-Roch theorem \eqref{RR} which states that $\ell_{16}=\ell_{22}=0$, and so there is no holomorphic differential to appear with these coefficients. On the other hand, $\ell_{40}=1$, and it is possible for $S_{40}$ to be non-zero. Using \eqref{Sp}, we calculate that in fact,
	\be
	S_{40}\approx7.0278\times10^{-4}.
	\ee
	This leads to a contradiction and hence shows that the bound $\tilde S^\infty_{12}$ cannot be saturated by $[0;2, 3, 7]$.
	
	Looking at the crossing equation \eqref{OPE Crossing} and the constraints~\eqref{eq:conditions2}, we can see that
	\begin{align}
		S_{12}-P^6_{\a^\infty}(0)\geq -\a^\infty_{28} S_{40}.
	\end{align}
	Since the bound is given by $\tl{S}^\infty_{12}=P^6_{\a^\infty}(0)$, we find that
	\be
	S_{12}-\tl{S}^\infty_{12}\geq -\a^\infty_{28} S_{40}, \label{Non-saturation 2}
	\ee
	which gives a lower bound on how far from saturation the bound we found is.
	
	Using the value of $\alpha^\infty_{28}$ and $S_{40}$ obtained above, we find the result~\eqref{OPE Non-saturation}
	\begin{align}\label{eq:finalS12nonsaturation}
		S_{12}-\tl{S}_{12}^\infty \gtrsim 4.1\times10^{-11}.
	\end{align}
	Since the value of $S_{12}$ on the $[0; 2, 3, 7]$ orbifold can be calculated \eqref{S12diff}, we have
	\begin{align}
		S_{12}-\tl{S}_{12}^{\Lambda=79} \approx 7.2\times10^{-11},
	\end{align}
	which implies that the bound $\tilde{S}_{12}^{\infty}$ cannot be much closer to saturation than the $\Lambda=79$ result.
	
	\subsection{Non-saturation of $\tl\l_1^\infty$ by the $[0; 2, 3, 7]$ Orbifold}
	\begin{figure}[t]
		\caption{}
		\begin{subfigure}[t]{0.49\textwidth}
			\includegraphics[width=\linewidth]{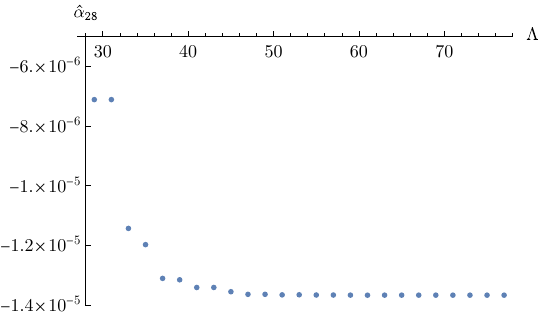}
			\caption{Plot showing the convergence of $\hat\alpha_{28}$ to a non-zero value with the single-correlator bound on $\l_1$ on with $n=6$.}
			\label{fig:237Alpha}
		\end{subfigure}
		\hfill
		\begin{subfigure}[t]{0.46\textwidth}
			\includegraphics[width=\linewidth]{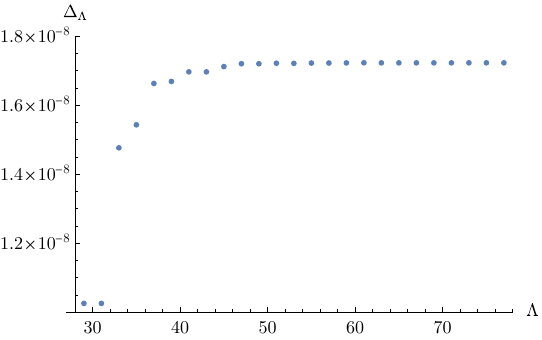}
			\caption{Plot of $\Delta_{\Lambda}$ with the single-correlator bound on $\l_1$ on with $n=6$.}
			\label{fig:237NonSat}
		\end{subfigure}
	\end{figure}
	
	The bound $\tl\l_1^{\Lambda=77}$ is almost saturated by the $[0; 2, 3, 7]$ orbifold. In order for the $\alpha_l$ to converge to an extremal functional, we need a different normalization to the normalization given in \eqref{eq:alphaconditions}. Namely, we can consider the functional given by $\hat{\alpha}_l=\frac{\alpha_l}{\alpha_1}$\footnote{As this is just rescaling \eqref{crossing0.5}, and $\alpha_1>0$, this functional still gives us a valid bound.}, and we do then see that $\hat\alpha_l$ seems to converge to $\hat\alpha_l^\infty$.
	
	Once again to show non-saturation we look for a term in \eqref{crossing0.5} that converges to a non-zero value, and once again, we find that $\alpha_{28}^\infty$ appears to be non-zero and we know from the previous section that $S_{40}$ on $[0; 2, 3, 7]$ is non-zero. Figure~\ref{fig:237Alpha} illustrates the convergence of $\hat{\alpha}_{28}$ explicitly. This already shows that this bound cannot be saturated either. In the rest of this section, we quantify how close to saturation it can be.
	
	We know that any valid $P^6_{\hat\alpha}$ can only be negative in the region $(-\infty, 0)\cup(0, \tl\l_1)$. In~\cite{Kravchuk:2021}, it was found using finite element methods that $\l_2\approx142.5551$ on the $[0; 2, 3, 7]$ orbifold, and so we can see that the only term in \eqref{crossing0.5} for this orbifold that can be negative is $P^6_{\hat\alpha}(\l_1)$ (as $\tl\l_1^\Lambda<46$ for all $\Lambda\geq1$). Hence, the crossing equation together with the positivity conditions from~\eqref{eq:alphaconditions} implies that
	\begin{align}
		\hat\alpha_{28} S_{40}>T_1P^6_{\hat\alpha}(\l_1).
	\end{align}
	
	Since for the functionals we obtain, $P^6_{\hat\alpha}$ is monotone increasing on the interval $[\l_1, \tl\l_1]$, we can invert this to find
	\begin{align}
		\l_1\leq(P^6_{\hat\alpha})^{-1}\left(\frac{\hat\alpha_{28} S_{40}}{T_1}\right).
	\end{align}
	Defining $\Delta_\Lambda=\tl\l_1^{\Lambda}- (P^6_{\hat\alpha})^{-1}\left(\frac{\hat\alpha_{28} S_{40}}{T_1}\right)$, this becomes
	\begin{align}
		\tl\l_1^\Lambda-\l_1\geq\Delta_\Lambda.
	\end{align}
	We calculate the value of $T_1$ on the $[0; 2, 3, 7]$ orbifold by numerically bootstrapping it with similar methods in Section \ref{setup}. We observe numerically that $\Delta_{\Lambda}$ strongly seems to be converging to a non-zero value, as is shown in Figure~\ref{fig:237NonSat}, and so on the $[0; 2, 3, 7]$ orbifold,
	\begin{align}
		\tl\l_1^\infty-\l_1\geq\Delta_\infty\approx1.7\times 10^{-8}.\label{237nonsat}
	\end{align}
	We do not know $\l_1$ on the $[0;2,3,7]$ orbifold with enough precision to evaluate the difference with the $\tl\l_1^{\L=77}$ bound given in~\eqref{eq:best237gap}. 
	
	\subsection{Non-saturation of $\tl\l_1^\infty$ by the Bolza Surface}
	\begin{figure}[t]
		\caption{}
		\begin{subfigure}[t]{0.47\textwidth}
			\includegraphics[width=\linewidth]{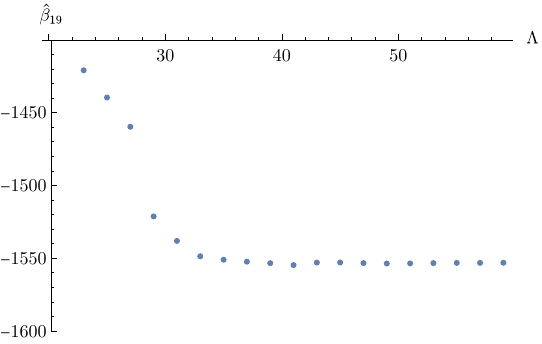}
			\caption{Plot showing the convergence of $\hat\beta_{19}$ to a non-zero value with the multi-correlator bound on $\l_1$ on genus-2 surfaces.}
			\label{fig:BolzaBeta}
		\end{subfigure}
		\hfill
		\begin{subfigure}[t]{0.48\textwidth}
			\includegraphics[width=\linewidth]{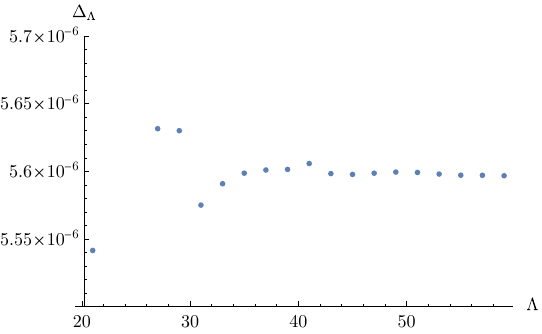}
			\caption{Plot of $\Delta_{\Lambda}$ with the multi-correlator bound on $\l_1$ on genus-2 surfaces.}
			\label{fig:BolzaNonSat}
		\end{subfigure}
	\end{figure}
	The bound on $\l_1$ for genus two surfaces \eqref{BolzaBound} is almost saturated by $\l_1$ on the Bolza surface, and if we rescale the functional such that $\hat\beta_1=1$, then we observe convergence to an extremal functional. We further know that on the Bolza surface, $\l_1$ is the only eigenvalue in $(0, \tl\l_1]$, as in~\cite{Aurich:1989vd}, they find that $\l_2\approx5.353$.
	By analysing the crossing equation \eqref{multi} with $n=1$, $\ell_n=2$, along with the conditions for the functional to produce a valid bound, we can see that
	\begin{align}
		\l_1\leq \begin{cases}
			F^{-1}\left(\frac{\hat\alpha_qS_{2+q}}{T_1}\right)&\text{if }q\text{ is even,}\\
			F^{-1}\left(\frac{\hat\beta_q A_{2+q}}{T_1}\right)&\text{if }q\text{ is odd,}\end{cases}\label{nonsaturationA}
	\end{align}
	where
	\begin{align}
		F(\l)&=P^1_{\hat\alpha}(\l)-P^1_{\hat\beta}(\l)-\frac{3}{2}P^1_{\hat\gamma}(\l)+\frac{1}{2}P^n_{\hat\delta}(\l),
	\end{align}
	and so we look for either a term $\hat\alpha^\infty_qS_{2n+q}$ or $\hat\beta^\infty_qA_{2n+q}$ that is non-zero.
	
	Once again looking at the coefficients, we see that $\hat\beta^\infty_{19}$ is non-zero from Figure~\ref{fig:BolzaBeta} and on the Bolza surface, by \eqref{Ap},
	\begin{align}
		A_{21}\approx1.8525\times10^{-9},
	\end{align}
	which is also non-zero on the Bolza surface. Therefore, the Bolza surface does not saturate $\tl\l_1^\infty$. Similarly to the last section, as $F$ is once again monotone increasing on the interval $[\l_1, \tl\l_1]$, we shall set $\Delta_\Lambda=\tl\l_1^{\Lambda}- F^{-1}\left(\frac{\hat\beta_{19}A_{21}}{T_1}\right)$, so that
	\begin{align}
		\tl\l_1^\Lambda-\l_1\geq\Delta_\Lambda.
	\end{align}
	Once again applying the numerical bootstrap to find the value of $T_1$, we calculate $\Delta_{\Lambda}$ on this surface, which strongly seems to be converging to a non-zero value, as is shown in Figure~\ref{fig:BolzaNonSat}, and so on the Bolza surface,
	\begin{align}
		\tl\l_1^\infty-\l_1\geq\Delta_\infty\approx 5.59683\times10^{-6},
	\end{align}
	giving us a bound on the non-saturation. In~\cite{Strohmaier:2011}, the value of $\l_1$ on the Bolza surface was found to high precision with finite element methods as
	\begin{align}
		\l_1=3.8388872588421995\dots,
	\end{align}
	and so comparing this to the bound \eqref{BolzaBound}, which has $\Lambda=59$, we can see that this is approximately $1.0389\times10^{-5}$ away from being saturated, and so $\tl\l_1^\infty$ cannot be much closer to saturation than $\tl\l_1^{\Lambda=59}$.
	
	\section{Conclusions\label{conclusions}}
	
	We have shown, by examining the functionals obtained from the bootstrap, that the bounds on $\lambda_1$, \eqref{general} and \eqref{BolzaBound} are not saturated by the $[0; 2, 3, 7]$ orbifold or the Bolza surface respectively, under reasonable assumptions about the convergence of the numerics. This was because of the fundamental obstruction that there are terms in the crossing equations \eqref{crossing0.5} and \eqref{multi}, which are converging to non-zero values. To show this obstruction, we related the OPE coefficients to Rankin-Cohen brackets and found ways to calculate the Petersson norms of these Rankin-Cohen brackets numerically on the $[0; 2, 3, 7]$ orbifold and on the Bolza surface.
	
	It is worth noting that although these bounds coming from simple systems of correlators are not saturated by either of these orbifolds, that does not mean that they cannot be improved by considering more sophisticated setups. For example, we only consider correlators of lowest-weight modular forms of level $\Gamma$, but there are obviously more sophisticated systems of correlators that could be formed using higher-weight modular forms, which might give tighter bounds.
	
	There are other applications of the bootstrap to spectral geometry, for example the spectrum of the Dirac operator on hyperbolic spin 2-orbifolds~\cite{Gesteau:2023brw}, and the spectrum of the Laplacian on closed Einstein manifolds~\cite{Bonifacio:2020xoc}, and closed hyperbolic 3-manifolds~\cite{Bonifacio:2023ban}. Based upon our results, we would expect that near-saturated bounds in these cases may also not actually be saturated, and the techniques we have used here could possibly be adapted to show non-saturation of these bounds as well. It would also be interesting to better understand the mechanism behind why these bounds are so close to saturation without actually being saturated.
	
	Better understanding this behaviour could also be useful from the point of view of near-saturated bootstrap bounds in CFTs. For example, we know that we have bootstrap bounds coming from a simple system of correlators that are almost saturated by the 3D critical Ising model, and an open question is whether these bounds are saturated by the 3D Ising model in the limit of an infinite number of derivatives acting on our crossing equations. Similarly to the methods we have used here, if we could show that there is a term in the crossing equations that is converging to a non-zero value, then that would show that the bounds are not saturated.
	
	\acknowledgments
	I would like to thank my supervisor Petr Kravchuk for so many enlightening conversations, and for much helpful advice about the direction of this paper. I would also like to thank King's College London for providing funding while this research was carried out. I would additionally like to thank Sridip Pal and Yixin Xu for numerical assistance, and to Dalimil Maz\'a\v{c} for helpful discussions and suggestions.
	

	\appendix
	
	\section{Uniformization Map and Metric for the $[0; 2, 3, 7]$ Orbifold and the Bolza Surface\label{appendix}}
	\subsection{$[0; 2, 3, 7]$ Orbifold}
	We recall from Section \ref{OPE Calculations}, that we have a biholomorphic isometry between a $[0; k_1, k_2, k_3]$ orbifold $X$, and the three-punctured sphere $\tl{X}$ with a specific metric that we will derive. This map can be lifted to a multi-valued holomorphic map $\tau$, from $\tilde{X}$ to $\mathbb{H}^2$. We call $\tau$ the uniformization map.
	
	We shall parameterize $\tl{X}$ as $\C\setminus\{0,1\} \simeq \mathbb{C}P^1\setminus \{0,1,\oo\}$, and shall place the orbifold point of order $k_1$ at $z_1=0$; the orbifold point of order $k_2$ at $z_2=1$; and the orbifold point of order $k_3$ at $z_3=\infty$.
	
	Away from the singularities, $\tau$ must be holomorphic, and in the neighbourhood of a singularity $z_i$, $\tau(z)=c_i+(z-z_i)^{1/k_i}+\dots$ for some constant $c_i$. Similarly, from the singularity at ${\infty}$, we get that at large $z$, $\tau(z^{-1})=c+z^{-1/k_i}+\dots$ for another constant $c$. Using~\eqref{Schwarzian} we can translate these constraints into conditions on the singularities of $\{\tau(z),z\}$. Recalling that $\{\tau(z),z\}$ is single-valued as the Schwarzian derivative is invariant under M\"obius transformations, one can then show that, with $\eta_i=\frac{k_i-1}{2k_i}$,
	\begin{align}
		\{\tau(z), z\}=&\frac{2\eta_1(1-\eta_1)}{z^2}+\frac{2\eta_2(1-\eta_2)}{(z-1)^2}\\
		&+(\eta_1(1-\eta_1)+\eta_2(1-\eta_2)-\eta_3(1-\eta_3))\left(\frac{1}{z}-\frac{1}{z-1}\right).
	\end{align}
	This is the starting point for deriving the hyperbolic metric for $\tilde{X}$. We omit this derivation and only quote the final result. For details, see~\cite{Hempel1988, Kravchuk:2021, Harlow:2011ny}.
	
	We define
	\begin{align}
		r&=\frac{\Gamma(2(1-\eta_1))^2\Gamma(1+\eta_1-\eta_2-\eta_3)\Gamma(\eta_1+\eta_2-\eta_3)\Gamma(\eta_1-\eta_2+\eta_3)\Gamma(\eta_1+\eta_2+\eta_3-1)}{\Gamma(2\eta_1)^2\Gamma(2-\eta_1-\eta_2-\eta_3)\Gamma(1-\eta_1+\eta_2-\eta_3)\Gamma(1-\eta_1-\eta_2+\eta_3)\Gamma(-\eta_1+\eta_2+\eta_3)},
	\end{align}
	and
	\begin{align}
		\tl{w}_1(z)&=z^{\eta_1}(1-z)^{\eta_2}{_2F_1}(\eta_1+\eta_2-\eta_3, \eta_1+\eta_2+\eta_3-1; 2\eta_1; z),\nn\\
		\tl{w}_2(z)&=(1-z)^{1-\eta_1}z^{\eta_2}{_2F_1}(1-\eta_1+\eta_2-\eta_3, -\eta_1+\eta_2+\eta_3; 2(1-\eta_1); z),
	\end{align}
	and the hyperbolic metric is then given by
	\begin{align}
		ds^2=e^{2\phi_{k_1k_2k_3}(z, \bar{z})}dz\,d\bar{z}=\frac{4r(1-2\eta_1)^2}{(r\tl{w}_1(z)\tl{w}_1(z)-\tl{w}_2(z)\tl{w}_2(z))^2}dz\,d\bar{z}.\label{metric}
	\end{align}
	The Petersson inner product \eqref{Petersson} can then be transformed onto $\tl{X}$, as
	\begin{align}
		\left\langle h_{n, a_1}(\tau), h_{n, a_2}(\tau)\right\rangle_{M_{2n}(\Gamma)}=\frac{1}{\vol (\tl{X})}\int_{\mathbb{C}}\,dz\,d\bar{z}\, e^{(2-2n)\phi_{k_1k_2k_3}(z, \bar{z})}\, \bar{\tl{h}_{n, a_1}(z)}\,\tl{h}_{n, a_2}(z),\label{norm}
	\end{align}
	where $\tl{h}_{n, a}(z)$ is the holomorphic differential corresponding to $h_{n, a}(\tau)$ according to \eqref{convert}. We shall write its induced norm as $\|\hat{h}_{n, a}(z)\,dz^n\|$.
	
	To apply this to our single correlator bootstrap with $n=6$, we need to find a holomorphic differential of degree 6 on $\tl{X}$. As $z$ is not a valid local coordinate at the orbifold points, we can have poles in $z$ here. For more details see~\cite{Kravchuk:2021}. The Riemann-Roch theorem \eqref{RR} tells us that there is only one such differential on the $[0; 2, 3, 7]$ orbifold (up to a multiplicative constant). This is given by
	\begin{align}
		f_6(z)dz^6=\frac{dz^6}{(z-1)^4 z^3}.
	\end{align}
	We can hence define a holomorphic differential of unit norm,
	\begin{align}
		\tl{h}_6(z)dz^6=\frac{f(z)dz^6}{\|f(z)dz^6\|}.\label{h6}
	\end{align}
	
	We can then see that the holomorphic differentials corresponding to \eqref{newprims}, are given by
	\begin{align}
		\hat{h}_{12+k}(z)dz^{12+k}&=\frac{2^k}{(23+k)_{k}}[\tl{h}_6(z)dz^6, \tl{h}_6(z)dz^6]_{D, k},
	\end{align}
	where the Rankin-Cohen bracket is given in \eqref{RCDeformed}. We can finally use \eqref{SpNormForm}, to see that the OPE coefficients in \eqref{crossing0} for the $[0; 2, 3, 7]$ orbifold are given by
	\begin{align}
		S_{2n+k}=\|\hat{h}_{12+k}(z)dz^{12+k}\|^2.\label{Sp}
	\end{align}
	As we have an explicit metric, this norm can be calculated by simply performing the numerical integration in \eqref{norm}, or by the techniques outlined in Appendix D.2 of \cite{Kravchuk:2021}.
	
	\subsection{Bolza Surface}
	\label{Bolza}
	The Bolza surface is defined as the one-point completion of the algebraic curve defined by~\cite{Bourque:2021}:
	\be
	\{(x, y)\in\mathbb{C}^2\,:\,y^2=x^5-x\}.
	\ee
	It is a closed genus-2 Riemann surface, and it can be viewed as the double cover of the sphere $\mathbb{C}P^1$ with six branch points. We shall use $x\in\mathbb{C}$ as the coordinate on the $\mathbb{C}P^1$, so that
	\be
	y=\pm\sqrt{x^5-x}.
	\ee
	We can then see that for a generic $x$, we have two values of $y$, and so the choice of square root determines which copy of the sphere we are on, however for $x=0, 1, i, -1, -i, \infty$, these two values of $y$ are equal, and so these are the branch points.
	
	Identifying the two copies of $\mathbb{C}P^1$, we get an orbifold at a particularly symmetric point in the $[0; 2, 2, 2, 2, 2, 2]$ moduli space. We can find a map $\psi$ from this orbifold onto the $[0; 4, 4, 4]$ orbifold by
	\be
	\psi(x)=\left(\frac{x^2+1}{x^2-1}\right)^2,
	\ee
	which maps the singularities to $0, 1, \infty$.
	
	We have the hyperbolic metric $\phi_{444}$ on this orbifold, and so we can pull this metric back onto the six-punctured spheres representing the Bolza orbifold as
	\be
	dx d\bar{x}e^{2\phi_{\text{Bolza}}(x, \bar{x})}=d\psi(x) d\bar{\psi(x)}e^{2\phi_{444}(\psi(x), \bar{\psi(x)})}.\label{bolza}
	\ee
	
	By similar logic to before, noting that we have six orbifold points of order 2 on each sphere, and considering the constraints this places on the uniformization map along with the constraint that $x\mapsto ix$, $y\mapsto \sqrt{i}y$ is a symmetry of this surface, we find that
	\begin{align}
		\{\tau(x), x\}=\frac{3}{8}\bigg(&\frac{1}{x^2}+\frac{1}{(x-1)^2}+\frac{1}{(x-i)^2}+\frac{1}{(x+1)^2}+\frac{1}{(x+i)^2}\nn\\
		&+\frac{-1}{x-1}+\frac{i}{x-i}+\frac{1}{x+1}+\frac{-i}{x+i}\bigg).\label{schwarzianBolza}
	\end{align}
	This allows us to calculate the Rankin-Cohen brackets on the Bolza surface with \eqref{RCDeformed}.
	
	On the Bolza surface, the Petersson norm \eqref{Petersson} of holomorphic differentials can be written as
	\begin{align}
		&\left\langle h_{n, a_1}(\tau), h_{n, a_2}(\tau)\right\rangle_{M_{2n}(\Gamma)}\\
		&=\frac{1}{\vol X}\Bigg(\int_{\mathbb{C}}\,dx\,d\bar{x}\, e^{(2-2n)\phi_{\text{Bolza}}(x, \bar{x})} \,\bar{\tl{h}_{n, a_1}\left(x, \sqrt{x^5-x}\right)}\,\tl{h}_{n, a_2}\left(x, \sqrt{x^5-x}\right)\nn\\
		&+\int_{\mathbb{C}}\,dx\,d\bar{x}\, e^{(2-2n)\phi_{\text{Bolza}}(x, \bar{x})} \,\bar{\tl{h}_{n, a_1}\left(x, -\sqrt{x^5-x}\right)}\,\tl{h}_{n, a_2}\left(x, -\sqrt{x^5-x}\right)\Bigg).
	\end{align}
	We shall once again write its induced norm as $\|\hat{h}_{n, a}(x, y)dx^n\|$.
	
	As $\ell_1=2$ by \eqref{RR}, we have a basis for holomorphic differentials of degree 1 on this surface by
	\begin{align}
		\tl{f}_{1, 1}(x, y)dx=\frac{dx}{y},
	\end{align}
	and
	\begin{align}
		\tl{f}_{1, 2}(x, y)dx=\frac{x\,dx}{y}.
	\end{align}
	
	From these, we use the Gram-Schmidt process to make an orthonormal basis $\tl{h}_{1, 1}(x, y)dx$, $\tl{h}_{1, 2}(x, y)dx$. If we then once again define the holomorphic differentials corresponding to \eqref{newprims},
	\begin{align}
		\hat{h}^{a_1, a_2}_{2+k}(x, y)dx^{2+k}&=\frac{2^k}{(3+k)_{k}}[\tl{h}_{1, a_1}(x, y)dx, \tl{h}_{1, a_2}(x, y)dx]_{D, k},
	\end{align}
	then according to \eqref{ApNormForm}, for $k$ odd,
	\begin{align}
		A_{2+k}&=\frac{1}{2}\sum_{a_1=1}^{2}\sum_{a_2=1}^{2}\left\|\hat{h}^{[a_1, a_2]}_{2+k}(x, y)dx^{2+k}\right\|^2\nn\\
		&=\left\|\hat{h}^{1, 2}_{2+k}(x, y)dx^{2+k}\right\|^2,\label{Ap}
	\end{align}
	which we can once again calculate by numerical integration.
	
	\bibliographystyle{JHEP} 
	\bibliography{bibliography} 
\end{document}